\documentclass[superscriptaddress,twocolumn,prd,aps,showpacs]{revtex4b3}
\usepackage{graphicx}
\usepackage{bm}
\begin{document}

\title{High energy photon-neutrino elastic scattering}
\author{Ali Abbasabadi}
\affiliation{Department of Physical Sciences,
Ferris State University, Big Rapids, Michigan 49307, USA}
\author{Alberto Devoto}
\affiliation{Dipartmento di Fisica, Universit\'a di Cagliari and
I.N.F.N., Sezione di Cagliari, Cagliari, Italy}
\author{Wayne W. Repko}
\affiliation{Department of Physics and Astronomy,
Michigan State University, East Lansing, Michigan 48824, USA}

\date{\today}
\begin{abstract}
\hspace*{0.1cm}
The one-loop helicity amplitudes for the elastic scattering process
$\gamma\nu\to\gamma\nu$ in the Standard Model are computed at high center of
mass energies. A general decomposition of the amplitudes is utilized to
investigate the validity of some of the key features of our results.  In the
center of mass, where $\sqrt{s} = 2\omega$, the cross section grows roughly as
$\omega^6$ to near the threshold for $W$-boson production, $\sqrt{s} = m_W$.
Although suppressed at low energies, we find that the elastic cross section
exceeds the cross section for $\gamma\nu\to\gamma\gamma\nu$ when
$\sqrt{s}>13\,$ GeV. We demonstrate that the scattered photons are circularly
polarized and the net value of the polarization is non-zero. Astrophysical
implications of high energy photon-neutrino scattering are discussed.
\end{abstract}
\maketitle

\vskip1pc

\section{Introduction}
\label{sec:1}

 The scattering process $\gamma\nu\to\gamma\nu$ has been studied in past
using the four-Fermi theory \cite{cm}, vector boson theories \cite{mjl,ls}, the
Standard Model \cite{cy,dr93}, and model-independent parameterizations
\cite{liu}. At low energies, $\omega \ll m_e$, where $\omega$ is the energy of
the photon in the center of mass and $m_e$ is the mass of the electron, the
cross section for the elastic scattering $\gamma\nu\to\gamma\nu$ is exceedingly
small \cite{dr93}. As it is shown in the Ref.\,\cite{dr97}, the cross section
for the inelastic process $\gamma\nu\to\gamma\gamma\nu$ is much larger than
that of the elastic scattering $\gamma\nu\to\gamma\nu$. However, with
increasing $\omega$, the elastic cross section eventually exceeds the
inelastic cross section. Here, we show that the cross over occurs at
$\omega\sim 7\,$ GeV.

Our explicit calculations, performed for high energies with massless electron
neutrinos, show that the scale for $\gamma\nu\to\gamma\nu$ scattering is set by
$m_W$. In fact, the cross section is effectively independent of the electron
mass, $m_e$ (except near forward and backward directions). Furthermore, since
the weak interaction violates parity, we find a large circular polarization of
the scattered photons.

In the next section, we utilize a decomposition of the elastic amplitude
${\cal A}(s,t,u)$ to obtain the general properties of and restrictions on the
helicity amplitudes ${\cal A}_{\lambda_1\lambda_2}(s,t,u)$. Section III
contains the numerical results for the complete one-loop helicity dependent
differential and total cross sections. This is followed by a discussion of the
production of circularly polarized photons in high energy
$\gamma\nu\to\gamma\nu$ collisions, and an estimate of the temperature at
which, during the thermal evolution of the universe, photons and neutrinos
decoupled.

\section{The $\bm\gamma\nu\to\gamma\nu$ elastic amplitude}
\label{sec:2}

The Lorentz-invariant amplitude for the process $\gamma\nu\to\gamma\nu$
can be expressed as \cite{addr99}
\begin{equation}\label{amp}
{\cal A}(s,t,u) = \;\bar{u}(p_2)\gamma_{\mu}(1 + \gamma_5)u(p_1)
{\cal M}_{\mu\alpha\beta}\,(\varepsilon_1)_{\alpha}(\varepsilon^*_2)_{\beta}\;,
\end{equation}
where $\varepsilon_1$ and $\varepsilon^*_2$ are the polarization vectors for
the incoming and outgoing photons, respectively, and the Mandelstam variables
$s$, $t$, and $u$ are defined by $s = -(p_1 + k_1)^2$, $t = -(p_1 - p_2)^2$,
and $u = -(p_1 - k_2)^2$. Here, the 4-momenta of the incoming neutrino and
photon are $p_1$ and $k_1$, respectively, with $p_2$ and $k_2$ denoting the
corresponding outgoing momenta. The tensor ${\cal M}_{\mu\alpha\beta}$ can be
expressed in terms of four linearly independent, gauge invariant tensors
$T^{(i)}_{\mu\alpha\beta},\;i = 1,\cdots ,4$\, \cite{addr99,nieves83}
\begin{eqnarray}\label{tenm}
{\cal M}_{\mu\alpha\beta} & =& {\cal M}_1(s,t,u)T^{(1)}_{\mu\alpha\beta} + {\cal
M}_2(s,t,u)T^{(2)}_{\mu\alpha\beta} \nonumber\\
&& + {\cal M}_3(s,t,u)T^{(3)}_{\mu\alpha\beta}
+ {\cal M}_4(s,t,u)T^{(4)}_{\mu\alpha\beta}\,.
\end{eqnarray}
To facilitate the inclusion of the Bose symmetry, which requires the invariance
of the amplitude under the exchange of the incoming and outgoing photons, we
use four tensors $T^{(i)}_{\mu\alpha\beta}$ which have a definite symmetry
under this exchange. For our choice, $T^{(1)}_{\mu\alpha\beta}$ is symmetric,
while $T^{(2)}_{\mu\alpha\beta}$, $T^{(3)}_{\mu\alpha\beta}$, and
$T^{(4)}_{\mu\alpha\beta}$ are antisymmetric. As a consequence, Bose symmetry
requires the following relations between the four scalar functions ${\cal
M}_i(s,t,u),\;i = 1,\cdots ,4$\,:
\begin{eqnarray}
{\cal M}_1(s,t,u) & = & {\cal M}_1(u,t,s)\,, \label{funm1} \\
{\cal M}_j(s,t,u) & = &\,-{\cal M}_j(u,t,s),\,j = 2,3,4\, \label{funm2}.
\end{eqnarray}

In the center of mass with massless neutrinos, the contractions of the tensors
$T^{(i)}_{\mu\alpha\beta}$ with the neutrino factor $\xi_{\mu} =
\bar{u}(p_2)\gamma_{\mu}(1 + \gamma_5)u(p_1)$ and polarization vectors
$(\varepsilon_1)_{\alpha}$ and $(\varepsilon^*_2)_{\beta}$, result in the
following helicity basis \cite{addr99}
\begin{eqnarray}
T^{(1)}_{\mu\alpha\beta}\,\xi_{\mu}(\varepsilon_1)_{\alpha}
(\varepsilon^*_2)_{\beta} & = &
-s\cos(\theta/2)[t(\lambda_1 +
\lambda_2 + 2\lambda_1\lambda_2) \nonumber\\
&& + 4s\lambda_1\lambda_2]\,,
\label{hel1}\\ [6pt]
T^{(2)}_{\mu\alpha\beta}\,\xi_{\mu}(\varepsilon_1)_{\alpha}
(\varepsilon^*_2)_{\beta} & = &
st\cos(\theta/2)(\lambda_1 - \lambda_2)\,, \label{hel2}\\ [6pt]
T^{(3)}_{\mu\alpha\beta}\,\xi_{\mu}(\varepsilon_1)_{\alpha}
(\varepsilon^*_2)_{\beta} & = &
st\cos(\theta/2)(1 - \lambda_1\lambda_2)\,, \label{hel3}\\ [6pt]
T^{(4)}_{\mu\alpha\beta}\,\xi_{\mu}(\varepsilon_1)_{\alpha}
(\varepsilon^*_2)_{\beta} & = &
-8\,\frac{s^2u}{t}\cos(\theta/2)\lambda_1\lambda_2\,,\label{hel4}
\end{eqnarray}
where $\theta$ is the angle between the incoming neutrino, which is moving in
the $+z$ direction, and the outgoing neutrino, $\lambda_1 = \pm 1$ is the
helicity of the incoming photon, and $\lambda_2 = \pm 1$ is the helicity of
the outgoing photon. Of these contractions, Eq.\,(\ref{hel2}) is antisymmetric
under the exchange of $\lambda_1$ and $\lambda_2$. The other three,
Eqs.\,(\ref{hel1}), (\ref{hel3}), and (\ref{hel4}), are symmetric under this
exchange. The imposition of time reversal invariance, which implies the
symmetry of the amplitude under the exchange of $\lambda_1$ and $\lambda_2$,
means that the T-violating part of the helicity basis, Eq.\,(\ref{hel2}), must
be excluded.

Using Eqs. (\ref{amp}), (\ref{tenm}), (\ref{hel1}), (\ref{hel3}), and
(\ref{hel4}), the helicity amplitudes ${\cal A}_{\lambda_1\lambda_2}(s,t,u)$
can be written as
\begin{eqnarray} \label{hamps}
{\cal A}_{++}(s,t,u) & = &
2s\cos(\theta/2)[2u{\cal M}_1(s,t,u) \nonumber\\
&& -4{su \over t}{\cal M}_4(s,t,u)]\,,\label{hamp1} \\
{\cal A}_{--}(s,t,u) & = &
2s\cos(\theta/2)[-2s{\cal M}_1(s,t,u) \nonumber\\
&& -4{su \over t}{\cal M}_4(s,t,u)]\,,\label{hamp2} \\
{\cal A}_{+-}(s,t,u) & = &
2s\cos(\theta/2)[(s-u){\cal M}_1(s,t,u) \nonumber\\
&& +t{\cal M}_3(s,t,u) +4{su \over t}{\cal M}_4(s,t,u)]\,,\label{hamp3} \\
{\cal A}_{-+}(s,t,u) & =  &{\cal A}_{+-}(s,t,u)\,,\label{hamp4}
\end{eqnarray}
where $t = -\frac{1}{2}s(1 - z)$, $u = -\frac{1}{2}s(1 + z)$, $z =
\cos\theta$, and we have assumed time reversal symmetry and, therefore,
omitted ${\cal M}_2(s,t,u)$. Notice that Eq.\,(\ref{hamp4}) is the result of
T-invariance, not Bose symmetry.

To include the requirements of Bose symmetry, Eqs.\,(\ref{funm1}) and
(\ref{funm2}), and also to exhibit the conservation of angular momentum in the
expressions for the helicity amplitudes, Eqs.\,(\ref{hamp1})--(\ref{hamp3}),
we define the following two independent scalar functions
\begin{eqnarray} \label{fg1}
{\cal F}(s,t,u) & = &
4{\cal M}_1(s,t,u) -8{s \over t}{\cal M}_4(s,t,u) \,,\label{f1} \\
{\cal G}(s,t,u) & = &
{\cal M}_3(s,t,u) -2{u \over t}{\cal M}_1(s,t,u)  \nonumber\\
&& +4{su \over t^2}{\cal M}_4(s,t,u)\,.\label{g1}
\end{eqnarray}
The interchange of $s$ and $u$ in the Eqs.\,(\ref{f1}) and (\ref{g1}),
together with the Bose symmetry requirements of Eqs.\,(\ref{funm1}) and
(\ref{funm2}), result in the following relations
\begin{eqnarray} \label{fg2}
{\cal F}(u,t,s) & = &
4{\cal M}_1(s,t,u) +8{u \over t}{\cal M}_4(s,t,u) \,,\label{f2} \\
{\cal G}(u,t,s) & = &
-{\cal M}_3(s,t,u) -2{s \over t}{\cal M}_1(s,t,u)  \nonumber\\
&& -4{su \over t^2}{\cal M}_4(s,t,u)\,.\label{g2}
\end{eqnarray}
Using Eqs.\,(\ref{f1})--(\ref{g2}) in Eqs.\,(\ref{hamp1})--(\ref{hamp3})
results in the following general expressions for the helicity amplitudes
\begin{eqnarray} \label{hampsfg}
{\cal A}_{++}(s,t,u) & = &
su\cos(\theta/2){\cal F}(s,t,u)\,,\label{hampfg1} \\
{\cal A}_{--}(s,t,u) & = &
-s^2\cos(\theta/2){\cal F}(u,t,s)\,,\label{hampfg2} \\
{\cal A}_{+-}(s,t,u) & = &
st\cos(\theta/2)[{\cal G}(s,t,u) - {\cal G}(u,t,s)]\,.\label{hampfg3}
\end{eqnarray}
In order to ensure the conservation of angular momentum in
Eqs.\,(\ref{hampfg1})--(\ref{hampfg3}), it is necessary to require that the
functions ${\cal F}(s,t,u)$, ${\cal F}(u,t,s)$, and $[{\cal G}(s,t,u) - {\cal
G}(u,t,s)]$ be non-singular in the limit $u \to 0$ (backward scattering). In
addition, the function $[{\cal G}(s,t,u) - {\cal G}(u,t,s)]$ must also be
non-singular in the limit $t \to 0$ (forward scattering).

From the Eqs.\,(\ref{hampfg1})--(\ref{hampfg3}), it is clear that the
interchange of $s$ and $u$, results in the following relation
\begin{eqnarray} \label{compact2}
{\cal A}_{\lambda_1\lambda_2}(s,t,u) & = &
{\cal A}_{{-\lambda_1}{-\lambda_2}}(u,t,s)\,,
\end{eqnarray}
where, under this interchange, the factor $s\cos(\theta/2) =
s\sqrt{\displaystyle {-u/s}}$ becomes $u\sqrt{\displaystyle {-s/u}} =
-s\sqrt{\displaystyle {-u/s}} = -s\cos(\theta/2)$.

We can use Eq.\,(\ref{compact2}) as a check of our calculation. To do this, we
must express the helicity amplitudes as functions of $s$ and $u$ such that
these functions remain well defined after the interchange $s\leftrightarrow
u$. A certain amount of care must be exercised when performing a numerical
check of Eq.\,(\ref{compact2}), because it is convenient to use the fact that
$s>0$ and $u<0$ when calculating ${\cal A}_{\lambda_1\lambda_2}(s,t,u)$. The
interchange of $s$ and $u$ would seem to move the numerical calculation into
the region $s<0$ and $u>0$ in order to make the comparison. However, no
additional calculation is required if one remembers that each of the diagrams
of Fig.\,1 has a counterpart in which the photons are interchanged. If a
particular scalar contribution to one of the direct diagrams is $f(s,u)$ the
corresponding crossed diagram will contribute $f(u,s)$. When it is assumed
that $s>0$ and $u<0$, the result is a function $f_1(s,u)$ for the direct
diagram and a different function $f_2(u,s)$ for the crossed diagram. The
function $f_1(x,y)$ is not defined when its first variable is negative and its
second positive, while $f_2(x,y)$ is not defined when its first variable is
positive and its second is negative. Since the $s\leftrightarrow u$
interchange essentially exchanges the direct and crossed contributions, we can
use
\begin{equation}\label{direct}
F(s,u)=\theta(s)\theta(-u)f_1(s,u)+\theta(-s)\theta(u)f_2(s,u)\,,
\end{equation}
for the direct contribution and
\begin{equation}\label{crossed}
F(u,s)=\theta(u)\theta(-s)f_1(u,s)+\theta(-u)\theta(s)f_2(u,s)\,.
\end{equation}
for the crossed contribution. Clearly, $F(s,u)$ and $F(u,s)$ are well defined
for both $s>0,u<0$ and $s<0,u>0$.

\section{Differential and total cross sections}

\begin{figure}
\centering
\includegraphics[width=3.25in]{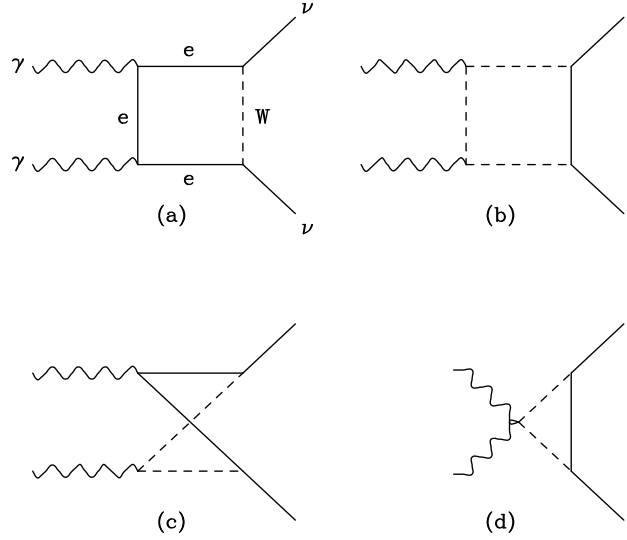}
\caption{Diagrams for the process $\gamma\nu_e \protect\to \gamma\nu_e$ are
shown. Diagram (d) gives a vanishing contribution. For each of (a), (b), (c)
there is also a diagram with the photons interchanged.} \label{diag}
\end{figure}

 We calculated the diagrams of Fig.\,\ref{diag} in a
nonlinear $R_{\xi}$ gauge such that the coupling between the photon, the
$W$-boson and the Goldstone boson vanishes \cite{dr93,gauge,pall}. Using
algebraic manipulation software {\sc{schoonschip}} \cite{schip} and
{\sc{form}} \cite{form}, the diagrams were decomposed in terms of scalar
n-point functions \cite{pv,tv}, and then checked numerically with the
{\sc{fortran}} codes {\sc{loop}} \cite{dk} and {\sc{ff}} \cite{ff}. In
addition, using Eqs.\,(\ref{direct}) and (\ref{crossed}), we demonstrated that
our numerical calculations for the helicity amplitudes, satisfy
Eqs.\,(\ref{hamp4}) and (\ref{compact2}). To simplify the calculations, we
assumed that $s, t, u \gg m_e^2$. Several of the scalar n-point functions
depended upon $\ln(m_e)$ or $\ln^2(m_e)$. However, it turns out that every
diagram of the Fig.\,{\ref{diag}} is independent of the $m_e$. Due to the
assumption $s, t, u \gg m_e^2$, in general, we do not expect that our results
be reliable near the forward and the backward directions, where $t \to 0$ and
$u \to 0$, respectively.

To study the degree of reliability, and also as a partial check of our
calculated helicity non-flip amplitudes near the forward direction, we use the
optical theorem, which relates the imaginary part of the non-flip amplitude
for elastic scattering in the forward direction to the total cross section for
the process $\gamma\nu\to W^+ e^-$ as
\begin{equation}\label{optical}
-{\frac{1}{s}}\;{\rm Im}\,{\cal A}_{\lambda\lambda}(\theta = 0) =
\sigma_{\lambda}\,.
\end{equation}
Here, $\lambda$ is the helicity of the photon, and $\sigma_{\lambda}$
represents the total cross section for a photon of given helicity $\lambda$,
after summation over the helicities of the W-boson and the electron. Our
explicit calculation gives
\begin{eqnarray}
\sigma_+ & = &
2\sqrt{2}G_F\alpha\left(\frac{\sqrt{\lambda(s,m_e^2,m_W^2)}}{s}(1+
\frac{m_W^2}{s})(1+2\frac{m_W^2}{s})\right. \nonumber \\
& &\left.+\frac{m_W^6}{s^3}\ell(s,m_W^2,m_e^2)\right. \nonumber \\
& &\left.-\frac{m_W^2}{s}(1-\frac{m_W^4}{s^2})\ell(s,m_e^2,m_W^2)\mbox{\rule{0pt}{16pt}}\right)\,,\\
[4pt] \sigma_- & =
&2\sqrt{2}G_F\alpha\left(\frac{\sqrt{\lambda(s,m_e^2,m_W^2)}}{s}(1-
\frac{m_W^2}{s})(1-2\frac{m_W^2}{s})\right. \nonumber \\
&
&\left.+\frac{m_W^2}{s}(1-\frac{m_W^2}{s})^2\ell(s,m_W^2,m_e^2)\right.\nonumber
\\
&  &\left.+\frac{m_W^2}{s}(1-\frac{m_W^2}{s})^2
\ell(s,m_e^2,m_W^2)\mbox{\rule{0pt}{16pt}}\right)\,,
\end{eqnarray}
where
\begin{eqnarray}
\lambda(x,y,z) & = &x^2+y^2+z^2-2xy-2xz-2yz\,, \\
\ell(x,y,z) & =
&\ln\left(\frac{x-y+z+\sqrt{\lambda(x,y,z})}{x-y+z-\sqrt{\lambda(x,y,z})}
\right)\,.
\end{eqnarray}
We retain the electron mass in the functions $\lambda(s,m_W^2,m_e^2)$,
$\ell(s,m_W^2,m_e^2)$ and $\ell(s,m_e^2,m_W^2)$ to ensure that the cross
sections vanish at threshold, $\sqrt{s}=m_W+m_e$. In the coefficients of these
functions, we have dropped powers of $m_e^2/m_W^2$. The spin averaged cross
section $(\sigma_+ + \sigma_-)/2$ agrees with the result previously obtained
by Seckel \cite{seck}.

Both cross sections approach the same constant when $\sqrt{s} \gg m_W$. Since
$\ell(s,m_W^2,m_e^2)\sim \ln(m_W^2/m_e^2)$, each cross section contains a
$\ln(m_e^2)$ term, which our numerical calculation of ${\cal
A}_{\lambda\lambda}(s,\theta)$ will not obtain in the forward direction due to
the assumption $s,t,u\gg m_e^2$. Nevertheless, we can use the optical theorem
to check the high energy behavior of our amplitudes because the
$\ln(m_W^2/m_e^2)$ terms in $\sigma_+$ and $\sigma_-$ depend very differently
on $m_W^2/s$; $(m_W^2/s)^3$ versus $m_W^2/s$. Indeed, for $\sqrt{s} \gtrsim
200$ GeV, and $\lambda = +1$, our calculations for the $-{\rm Im}\,{\cal
A}_{++}(\theta = 0)/s$ and $\sigma_{+}$ are in good agreement with
Eq.\,(\ref{optical}), while for $\lambda = -1$, the cross section $\sigma_{-}$
is almost a factor of two larger than the quantity $-{\rm Im}\,{\cal
A}_{--}(\theta = 0)/s$. However, at higher energies, we find an excellent
agreement with Eq.\,(\ref{optical}) for both $\lambda = \pm 1$ helicities.

Before leaving the subject of the behavior of the non-flip forward helicity
amplitudes, it is worth noting that the exact value of the forward helicity
amplitude ${\cal A}_{\lambda\lambda}(s)$ can be obtained using the dispersion
relation
\begin{equation}
{\cal
A}_{\lambda\lambda}(s)=\frac{s^2}{\pi}\int_{(m_W+m_e)^2}^{\infty}\frac{ds'}{s'}
\left(\frac{\sigma_{\lambda}(s')}{s'-s}+\frac{\bar{\sigma}_{\lambda}(s')}{s'+s}\right),
\end{equation}
where $\bar{\sigma}_{\lambda}(s)$ is the cross section in the channel
$\gamma\bar{\nu}\to W^-e^+$. It is not difficult to show that
$\bar{\sigma}_{\lambda}(s) = \sigma_{-\lambda}(s)$, in which case we have
\begin{equation}\label{disp}
{\cal
A}_{\lambda\lambda}(s)=\frac{s^2}{\pi}\int_{(m_W+m_e)^2}^{\infty}\frac{ds'}{s'}
\left(\frac{\sigma_{\lambda}(s')}{s'-s}+\frac{\sigma_{-\lambda}(s')}{s'+s}\right).
\end{equation}
This expression obeys the symmetry relation Eq.\,(\ref{compact2}) specialized
to the forward direction $t=0$,
\begin{equation}
{\cal A}_{\lambda\lambda}(s)={\cal A}_{-\lambda -\lambda}(-s).
\end{equation}
To leading order in $s$, the dispersion integral can be evaluated and gives
\begin{eqnarray}
A_{++}(s)& =& A_{--}(s)\nonumber \\
& = &\frac{\sqrt{2}G_F\alpha s^2}{\pi m_W^2}
\left[\frac{2}{3}\ln\left(\frac{m_W^2}{m_e^2}\right)+\frac{1}{2}\right]\,,
\end{eqnarray}
in agreement with the low energy result of \cite{dr93}. Details will be
presented elsewhere.

\begin{figure}[!]
\centering
\includegraphics[width=3.25in]{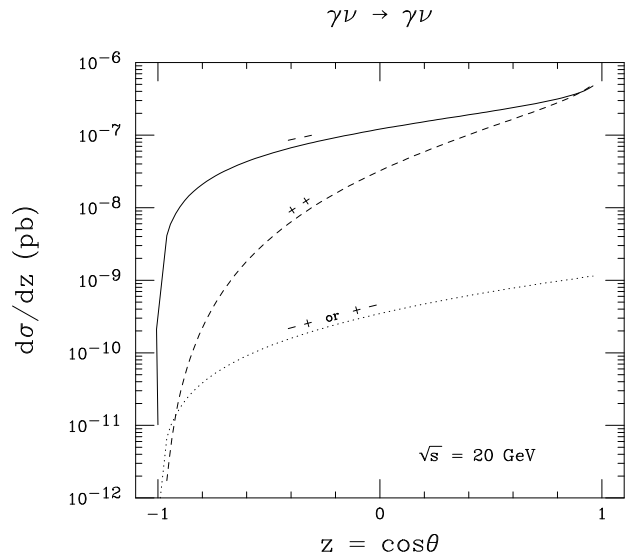}
\caption{The helicity dependent differential cross sections for
$\gamma\nu\protect\to\gamma\nu$ are shown for $\protect\sqrt{s} = 20$\,GeV.
The solid line is $d\sigma_{--}/dz$, the dashed line is $d\sigma_{++}/dz$, and
the dotted line is $d\sigma_{+-}/dz=d\sigma_{-+}/dz$.} \label{dsigma_20}
\end{figure}

\begin{figure}
\centering
\includegraphics[width=3.25in]{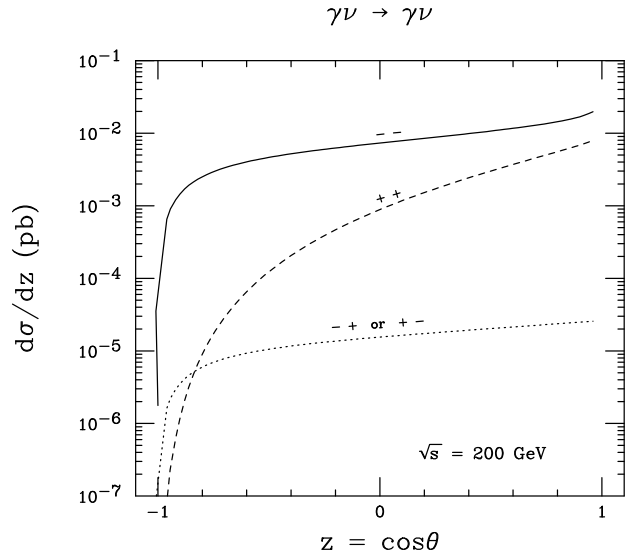}
\caption{Same as Fig.\,\ref{dsigma_20} with
$\protect\sqrt{s} = 200$\,GeV.}\label{dsigma_200}
\end{figure}

In Figs.\,\ref{dsigma_20} and \ref{dsigma_200} we show the differential cross
sections for elastic scattering using
\begin{equation}
\frac{d\sigma_{\lambda_1\lambda_2}}{dz} = \frac{1}{32\pi s}|{\cal
A}_{\lambda_1\lambda_2}|^2\,,
\end{equation}
where, $\lambda_1$ and $\lambda_2$ are the helicities of the incoming and
outgoing photons, respectively. Figs.\,\ref{dsigma_20} and \ref{dsigma_200}
illustrate the identity of the amplitudes ${\cal A}_{+-}$ and ${\cal A}_{-+}$,
as required by Eq.\,(\ref{hamp4}). They also show the vanishing of the
amplitudes for backward scattering, $\theta = \pi$. However, they do not
exhibit the vanishing of the ${\cal A}_{+-}$ or ${\cal A}_{-+}$ in the forward
direction, $\theta = 0$. As we stated earlier we do not, in general, expect our
results be reliable near the forward ($t \to 0$) and backward ($u \to 0$)
directions on a scale $\sim m_e^2$. In our calculations, we have, for
instance, replaced factors such as $t/(t-m_e^2)$ with $1$. This situation is
similar to that encountered when calculating the amplitudes for quark+gluon
$\to$ quark+photon \cite{dprk}, where the existence of the kinematic zeros
proportional to $t$ in the forward direction is obscured by the assumption
that the quarks are massless.

\begin{figure}
\centering
\includegraphics[width=3.25in]{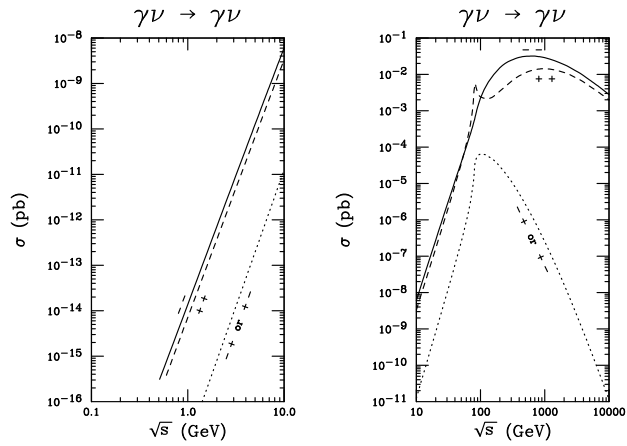}
\caption{The helicity dependent total cross sections for
$\gamma\nu\protect\to\gamma\nu$ are shown. The solid line is $\sigma_{--}$,
the dashed line is $\sigma_{++}$, and the dotted line is $\sigma_{+-}=
\sigma_{-+}$.} \label{sigma}
\end{figure}

 The total cross sections, for helicities $\lambda_1$
and $\lambda_2$, are given by
\begin{equation}
\sigma_{\lambda_1\lambda_2} = \frac{1}{32\pi s}\int_{-1}^1dz
|{\cal A}_{\lambda_1\lambda_2}|^2\,,
\end{equation}
and are plotted in Fig.\,\ref{sigma}. Shown in dots is the cross section for
helicity flip scattering, which can be seen to be much smaller than the cross
sections for the helicity non-flip scattering. This feature seems not to be a
consequence of any symmetry, but it is reminiscent of the low energy case
($\sqrt{s} \ll m_e$), where the helicity flip amplitudes vanish \cite{dr93}. In
Fig.\,\ref{sigma_avg}, we show the total cross section for an unpolarized
initial photon,
\begin{equation}
\sigma_{\gamma\nu\to\gamma\nu} = \frac{1}{2}\left(
\sigma_{--} + \sigma_{++} + \sigma_{+-} + \sigma_{-+}\right)\,.
\end{equation}
This figure illustrates the roughly $s^3$ behavior of the total cross section
to near the threshold for the $W$-boson production. A fit to the points in
Fig.\,\ref{sigma_avg}, for $m_e \ll \omega \ll m_W$, yields
\begin{equation}\label{siggngn}
\sigma_{\gamma\nu\to\gamma\nu} = 6.7\times 10^{-33}
\left(\frac{\omega}{m_e}\right)^6\,\mbox{\rm pb}\,,
\end{equation}
where $\omega = \sqrt{s}/2$ is the energy of a photon (or a neutrino). In the
Ref.\,\cite{dkr98}, it is shown that the cross section for the process
$\gamma\nu\to\gamma\gamma\nu$, in the range of energies $m_e \ll \omega \ll
m_W$, can be written as
\begin{equation}\label{siggnggn}
\sigma_{\gamma\nu\to\gamma\gamma\nu} = 1.74\times 10^{-16}
\left(\frac{\omega}{m_e}\right)^2\,\mbox{\rm pb}\,.
\end{equation}
Comparison of the Eq.\,(\ref{siggngn}) with the
Eq.\,(\ref{siggnggn}), shows that the two cross sections are equal
for $\omega = 1.27 \times 10^{4}m_e$ or $\sqrt{s} = 13$ GeV.
Therefore, at sufficiently high energies, the process
$\gamma\nu\to\gamma\nu$ dominates the process
$\gamma\nu\to\gamma\gamma\nu$.

\section{Discussion and conclusions}

As shown above, the cross section for elastic scattering is larger than the
cross section for the inelastic scattering when $\sqrt{s}\agt 13\,$ GeV. Since
the weak interaction violates parity, the final photons in both processes
acquire circular polarization. In the case of inelastic scattering, it is
shown in the Refs.\,\cite{dr97} and \cite{dkr98} that the circular
polarization is of order $20-30$\% for center of mass energies less than
$100m_e$.

\begin{figure}
\centering
\includegraphics[width=3.25in]{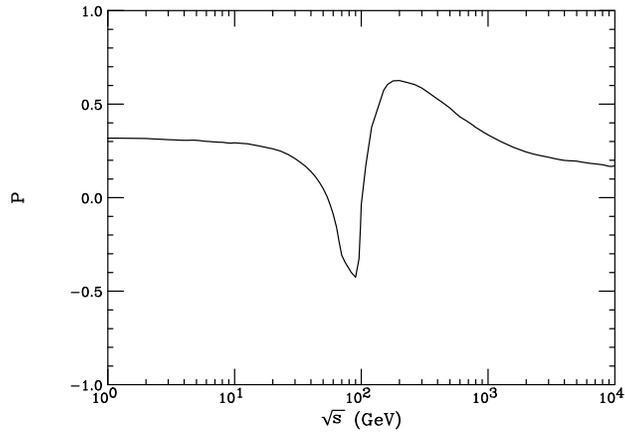}
\caption{The circular polarization P of the final photon in the process
$\gamma\nu\to\gamma\nu$, as defined in the Eq.\,(\ref{p1}), is shown. For
$\sqrt{s} \ll 2m_e$ \cite{dr93}, P is $1/3$.} \label{polarization_energy}
\end{figure}

To assess the degree of circular polarization of the final photon in the
elastic scattering, we define the polarization P as
\begin{equation}\label{p1}
\rm P = \frac{\sigma_{--} + \sigma_{+-} - \sigma_{-+} - \sigma_{++}}
{\sigma_{--} + \sigma_{+-} + \sigma_{-+} + \sigma_{++}}\,,
\end{equation}
or, since $\sigma_{+-} = \sigma_{-+} \ll \sigma_{--}$,
\begin{equation}\label{p2}
\rm P \simeq \frac{\sigma_{--} - \sigma_{++}}
{\sigma_{--} + \sigma_{++}}\,.
\end{equation}
In Fig.\,\ref{polarization_energy}, we have plotted the polarization P as a
function of the center of mass energy $\sqrt{s}$. As is clear from this graph,
for wide range of energies above $1$ GeV, the polarization is of order
$20-30$\%, while for energies around $200$ GeV it can reach over $60$\%. To
find P for center of mass energies $\sqrt{s} \ll 2m_e$, we can use the helicity
amplitudes in Eq.\,(4) of Ref.\,\cite{dr93}, to obtain ${\rm P} = 1/3$.

\begin{figure}
\centering
\includegraphics[width=3.25in]{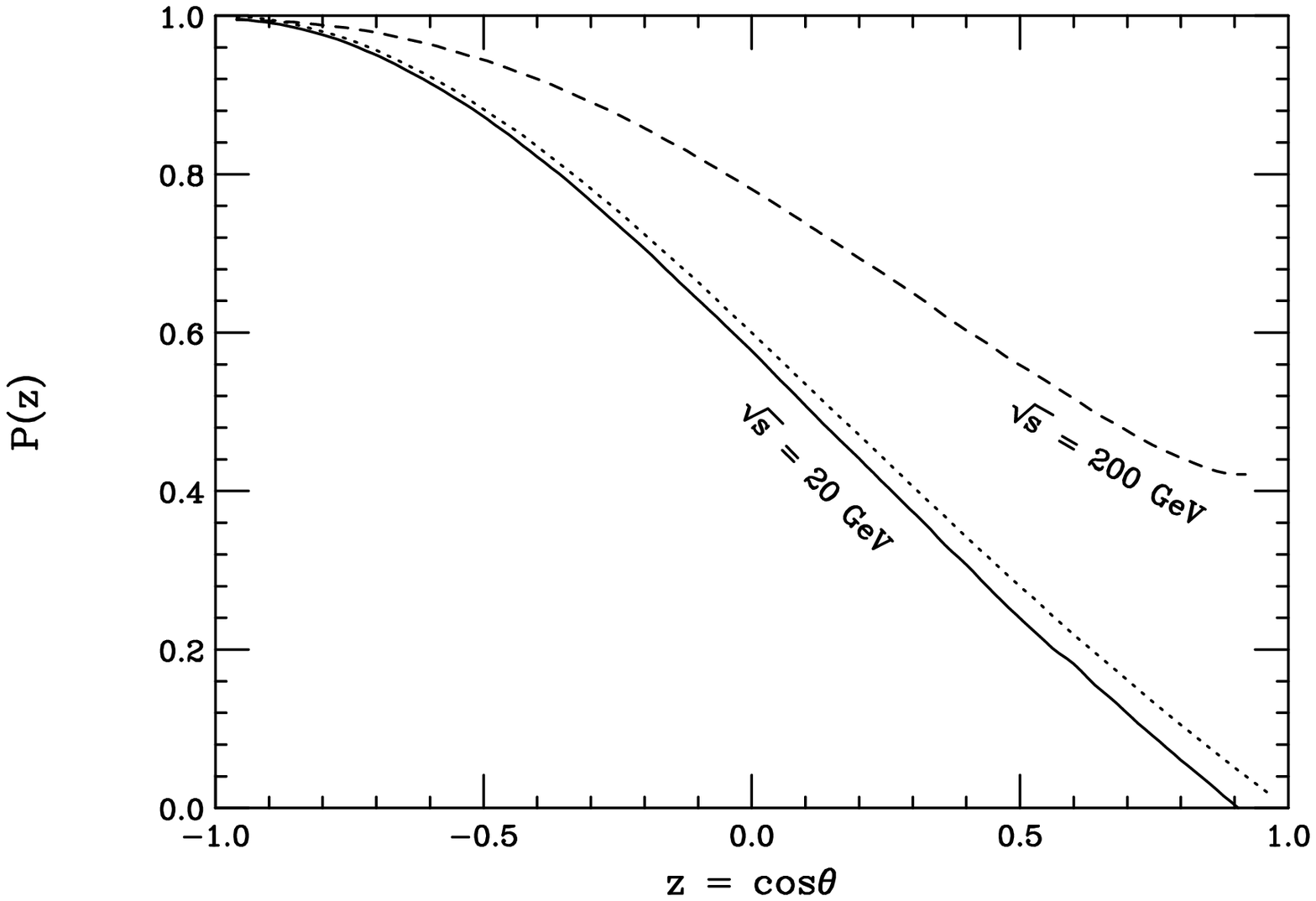}
\caption{The circular polarization ${\rm P}(z)$ of the final photon in the
process $\gamma\nu\to\gamma\nu$, as defined in the Eq.\,(\ref{p4}), is shown.
The solid and the dashed lines are polarization for the center of mass
energies of $20$ GeV and $200$ GeV, respectively, while the dotted line is for
$\sqrt{s}\ll 2m_e$, which is taken from the Ref.\,\cite{dr93}.}
\label{polarization_angle}
\end{figure}

The angular dependence of the the final photon's polarization can be obtained
from ${\rm P}(z)$ defined as
\begin{equation}\label{p3}
{\rm P}(z) = \frac{{d\sigma_{--}}/dz + d\sigma_{+-}/dz - d\sigma_{-+} /dz -
d\sigma_{++} /dz} {d\sigma_{--} /dz + d\sigma_{+-} /dz + d\sigma_{-+} /dz +
d\sigma_{++}/dz}\,,
\end{equation}
or, since $d\sigma_{+-}/dz = d\sigma_{-+}/dz$,
\begin{equation}\label{p4}
{\rm P}(z) = \frac{d\sigma_{--} /dz - d\sigma_{++} /dz} {d\sigma_{--} /dz +
2d\sigma_{+-} /dz +d\sigma_{++} /dz}\,.
\end{equation}
The polarization ${\rm P}(z)$ is plotted in Fig.\,\ref{polarization_angle} as a
function of $z = \cos\theta$. In this figure, the solid line is effectively
unchanged for the range of center of mass energies $1\,{\rm GeV} \lesssim
\sqrt{s} \lesssim 30\, {\rm GeV}$. The dashed curve shows that the forward
amplitudes ${\cal A}_{++}(s)$ and ${\cal A}_{--}(s)$ are not equal above the
threshold for $W$ production. This is consistent with the dispersion relation
Eq.\,(\ref{disp}). Also included in this figure is the polarization for the
case $\sqrt{s} \ll 2m_e$, which is based on the Eq.\,(5) of the
Ref.\,{\cite{dr93}} (${\rm P}(z) = - {\cal P}(\theta)$).

It is worth noting that both the $s^3$ behavior of the elastic cross section
and the angular dependence of $\mathrm{P}(z)$ obtained in the low energy limit
\cite{dr93} persist to energies of order $m_W$. This means that a low energy
effective interaction of the form \cite{dkr}
\begin{equation}
{\cal L}_{\rm{eff}} =
\frac{1}{8\pi}\frac{g^2\alpha}{m_W^4}AT^{\nu}_{\lambda\rho}T^{\gamma}_{\lambda\rho}\,,
\end{equation}
where $T^{\nu}_{\lambda\rho}$ and $T^{\gamma}_{\lambda\rho}$ are the
symmetrical energy-momentum tensors of the neutrino and the photon, gives an
accurate description of elastic scattering to quite high energies.

\begin{figure}
\centering
\includegraphics[width=3.25in]{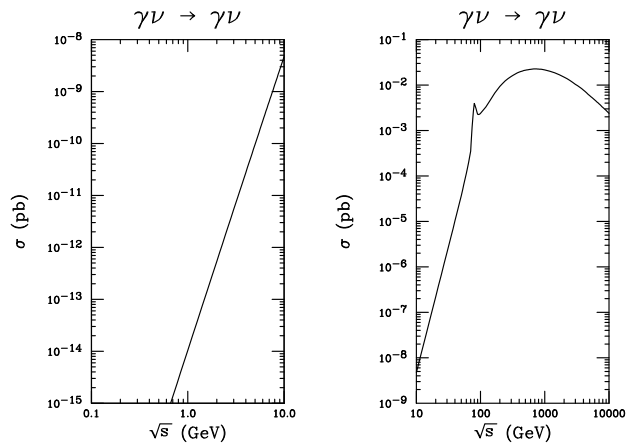}
\caption{The total cross section $\sigma_{\gamma\nu\to\gamma\nu}$ for
unpolarized photons is shown.} \label{sigma_avg}
\end{figure}

Finally, to investigate the importance of the reaction $\gamma\nu\to\gamma\nu$
in cosmology, we define $\sigma_\nu$ by
\begin{equation}\label{relic1}
\sigma_\nu = \frac{1}{n_\nu c\, t}\,,
\end{equation}
where $n_\nu$ is the neutrino number density (number of neutrinos per unit
volume), $c$ is the speed of light, and $t$ is the expansion time of the
universe. Taking $n_\nu = 56\, {\rm cm}^{-3}$ and $t = 15 \times 10^9$ years,
we find the present value of $\sigma_\nu$ to be
\begin{equation}\label{relic2}
\sigma_\nu = 1.26 \times 10^6\,\mbox{\rm pb}\,.
\end{equation}
The mean number of collisions between a photon and relic neutrinos \cite{p93}
is of order $\sigma_{\gamma\nu\to\gamma\nu}$/$\sigma_\nu$. From the
Fig.\,\ref{sigma_avg} and Eq.\,(\ref{relic2}) we find that, regardless of the
center of mass energy, $\sigma_{\gamma\nu\to\gamma\nu}$/$\sigma_\nu \lesssim
10^{-7}$. Therefore, the process $\gamma\nu\to\gamma\nu$ effectively ceased to
occur early in the evolution of the universe.

To estimate the time (or the temperature) at which the decoupling of photons
and neutrinos in the process $\gamma\nu\to\gamma\nu$ took place, we must
determine when the value of the ratio
$\sigma_{\gamma\nu\to\gamma\nu}$/$\sigma_\nu$ is of order 1,
\begin{equation}\label{rate}
\sigma_{\gamma\nu\to\gamma\nu}/\sigma_\nu \sim 1 \,.
\end{equation}
To do this, consider the product $\sigma(p_\gamma,p_\nu)\,v_{\gamma\nu}$, where
$\sigma(p_\gamma,p_\nu)$ is the cross section for the scattering of a photon
of momentum $\vec{p}_\gamma$ with a neutrino of momentum $\vec{p}_\nu$, and
$v_{\gamma\nu}$ is the magnitude of their relative velocity. We define the
average value of this product by
\begin{equation}\label{sigvavg1}
\langle\sigma(p_\gamma,p_\nu)\, v_{\gamma\nu}\rangle =
\frac {\int dn_{\gamma}\,dn_{\nu}\,
\sigma(p_\gamma,p_\nu)\,v_{\gamma\nu}}
{\int dn_{\gamma}\,dn_{\nu}}\,,
\end{equation}
where
\begin{eqnarray}
dn_{\gamma} = \frac{1}{(2\pi)^3}
\frac{2d^{3}p_\gamma}{e^{E_{\gamma}/T} - 1}
\,,\label{dngamma}\\
dn_{\nu} = \frac{1}{(2\pi)^3}
\frac{d^{3}p_\nu}{e^{E_{\nu}/T} + 1}\,.\label{dnnu}
\end{eqnarray}
Here, $E_\gamma$ and $E_\nu$ are the energies of the photon and
neutrino, respectively, and the factor 2 in the Eq.\, (\ref{dngamma})
is due to the number of helicity states of the photon. After performing
the integration in the denominator of the Eq.\,(\ref{sigvavg1}),
we obtain
\begin{equation}\label{sigvavg2}
\langle\sigma(p_\gamma,p_\nu)\, v_{\gamma\nu}\rangle =
\frac{1}{n_{\gamma}n_{\nu}}\int dn_{\gamma}dn_{\nu}\,
\sigma(p_\gamma,p_\nu)\,v_{\gamma\nu}\,,
\end{equation}
where $n_{\gamma}$ and $n_{\nu}$, the number
densities for the photons and neutrinos, respectively, are
\begin{eqnarray}
n_{\gamma} = \frac{1}{(2\pi)^3}
\int \frac{2d^{3}p_\gamma}{e^{E_{\gamma}/T} - 1} =
\frac{2\zeta (3)T^3}{\pi ^2}
\,,\label{ngamma}\\
n_{\nu} = \frac{1}{(2\pi)^3} \int \frac{d^{3}p_\nu}{e^{E_{\nu}/T} + 1} =
\frac{3\zeta (3)T^3}{4\pi ^2} \,.\label{nnu}
\end{eqnarray}

Using the invariance of
$\sigma(p_\gamma,p_\nu)\,v_{\gamma\nu}\,E_\gamma\,E_\nu$, the product
$\sigma(p_\gamma,p_\nu)\,v_{\gamma\nu}$ can be expressed in terms of the cross
section $\sigma_{\rm CM}$ in the center of mass as
\begin{equation}\label{sigv1}
\sigma(p_\gamma,p_\nu)\,v_{\gamma\nu} =
\sigma_{\rm CM} \frac{2E_{\rm CM}^2}{E_{\gamma}E_{\nu}}\,,
\end{equation}
where $\sigma_{\rm CM} = \sigma_{\gamma\nu\to\gamma\nu}$ is given by
Eq.\,(\ref{siggngn}) with $\omega = E_{\rm CM}$. The center of mass energy
$E_{\rm CM}$ for a photon (or a neutrino) in terms of $E_\gamma$, $E_\nu$, and
$\theta_{\gamma\nu}$, the angle between $\vec{p}_\gamma$ and $\vec{p}_\nu$, is
\begin{equation}\label{ecm}
E_{\rm CM} = \sqrt{E_{\gamma}E_{\nu}} \sin (\theta_{\gamma\nu}/2)\,.
\end{equation}
Therefore, Eqs.\,(\ref{siggngn}), (\ref{sigv1}), and (\ref{ecm}) give
\begin{equation}\label{sigv2}
\sigma(p_\gamma,p_\nu)\,v_{\gamma\nu} = 6.7 \times 10^{-33}\,
\frac{2E_{\gamma}^3 E_{\nu}^3}{m_e^6}\, \sin^8
(\theta_{\gamma\nu}/2)\,\,\,\mbox{\rm pb}\,.
\end{equation}
Using this result in Eq.\,(\ref{sigvavg2}) and performing the integration, we
find
\begin{equation}\label{sigvavg3}
\langle\sigma(p_\gamma,p_\nu)\, v_{\gamma\nu}\rangle =
6.7 \times 10^{-33}\,
\frac{124}{59535}\frac{\pi^{12}}{[\zeta(3)]^2}\frac{T^6}{m_e^6}
\,\,\,\mbox{\rm pb}\,.
\end{equation}

 The average number of collisions that a photon makes with
neutrinos through the process $\gamma\nu\to\gamma\nu$ during the time $t$ is
\begin{equation}\label{ngammaavg}
{\cal N}_\gamma =
\langle\sigma(p_\gamma,p_\nu)\, v_{\gamma\nu}\rangle \,n_\nu \,t\,.
\end{equation}
Similarly, the average number of collisions that a neutrino makes with photons
through the process $\gamma\nu\to\gamma\nu$ during the time $t$ is
\begin{equation}\label{nnuavg1}
{\cal N}_\nu =
\langle\sigma(p_\gamma,p_\nu)\, v_{\gamma\nu}\rangle \,n_\gamma \,t\,.
\end{equation}
For the duration time, we take an expansion time, which is given by \cite{w72}
\begin{equation}\label{expansiont}
t = \frac{2 \times 10^{20}}{T^2}\;\,\mbox{\rm s}\,,
\end{equation}
where $T$ is the temperature in Kelvin. To estimate the temperature at which
the thermal decoupling of the photons and neutrinos took place, we use the
criterion
\begin{equation}\label{criterion1}
{\rm max}\left [{\cal N}_\gamma , {\cal N}_\nu \right] \sim 1 \,.
\end{equation}
It is clear from Eqs.\,(\ref{ngamma}) and (\ref{nnu}) that
$n_\gamma > n_\nu$, and from Eqs.\,(\ref{ngammaavg}) and
(\ref{nnuavg1}) that ${\cal N}_\gamma < {\cal N}_\nu $. Therefore,
Eq.\,(\ref{criterion1}) gives
\begin{equation}\label{criterion2}
{\cal N}_\nu \sim 1 \,.
\end{equation}
Using Eqs.\,(\ref{expansiont}), (\ref{ngamma}), (\ref{sigvavg3}), and
(\ref{nnuavg1}), we find
\begin{equation}\label{nnuavg2}
{\cal N}_\nu = 2.5 \times 10^{-92}\, T^7\,.
\end{equation}
Thus, from Eq.\,(\ref{criterion2}) we find the decoupling temperature $T \sim
1.2 \times 10^{13}$ K, or $T \sim 1$ GeV. This translates into an expansion
time of about $1.4\times 10^{-6}$ s. The temperature $T \sim 1$ GeV
corresponds to a center of mass energy $\sqrt{s}\sim 2$ GeV, which is well
within the range of validity of Eq.\,(\ref{siggngn}).

\acknowledgments We wish to thank Duane Dicus for numerous helpful
discussions. One of us (A.A.) wishes to thank the Department of Physics and
Astronomy at Michigan State University for its hospitality and computer
resources. This work was supported in part by the National Science Foundation
under Grants No. PHY-9802439 and PHY-0070443 and by Confinziamento MURST-PRIN.

\onecolumn

\vspace{.10in}

\begin{thebibliography}{00}
\bibitem{cm} H.-Y. Chiu and P. Morrison,
Phys. Rev. Lett. {\bf 5}, 573 (1960).
\bibitem{mjl} M. J. Levine, Nuovo Cimento {\bf 48A}, 67 (1967).
\bibitem{ls} L. F. Landovitz and W. M. Schreiber,
Nuovo Cimento {\bf 2A}, 359 (1971).
\bibitem{cy} V. K. Cung and M. Yoshimura,
Nuovo Cimento {\bf 29A}, 557 (1975).
\bibitem{dr93} D. A. Dicus and W. W. Repko,
Phys. Rev. D {\bf 48}, 5106 (1993).
\bibitem{liu} J. Liu, Phys. Rev. D {\bf 44}, 2879 (1991).
\bibitem{dr97} D. A. Dicus and W. W. Repko,
Phys. Rev. Lett. {\bf 79}, 569 (1997).
\bibitem{addr99} A. Abbasabadi, A. Devoto, D. A. Dicus,
and W. W. Repko, Phys. Rev. D {\bf 59}, 013012 (1999).
\bibitem{nieves83} For a related decomposition in the
case ${\nu}^\prime \to \nu\gamma\gamma$, see J. F. Nieves, Phys. Rev. D {\bf
28}, 1664 (1983).
\bibitem{gauge}  M. B. Gavela, G. Girardi, C. Malleville, and P. Sorba,
Nucl. Phys. {\bf B193}, 257 (1981); M. Bace and N. D. Hari Dass,
Ann. of Phys. {\bf 94}, 349 (1975).
\bibitem{pall} J. F. Nieves, P. B. Pal, and D. G. Unger,
Phys. Rev. D {\bf 28}, 908 (1983).
\bibitem{schip} M. J. G. Veltman, ``{\sc{schoonschip}} A Program for Symbol
Handling," University of Michigan, report, 1984 (unpublished).
\bibitem{form} J. A. M. Vermaseren, ``The Symbolic Manipulation Program
{\sc{form}}," Report No. KEK-TH-326, 1992 (unpublished).
\bibitem{pv} G. Passarino and M. Veltman,
Nucl. Phys. {\bf B160}, 151 (1979).
\bibitem{tv} G. 't Hooft and M. Veltman,
Nucl. Phys. {\bf B153}, 365 (1979).
\bibitem{dk} C. Kao and D. A. Dicus, {\sc{loop}}, a {\sc{fortran}}
program for evaluating loop integrals based on the results in
Refs.\,{\cite{pv} and \cite{tv}}.
\bibitem{ff} G. J. van Oldenborgh, NIKHEF-H/90-15, (1990).
\bibitem{seck}D. Seckel, Phys. Rev. Lett. {\bf 80}, 900 (1998).
\bibitem{dprk} A. Devoto, J. Pumplin, W. Repko, and G. L. Kane,
Phys. Rev. Lett. {\bf 43}, 1062 (1979).
\bibitem{dkr98} D. A. Dicus, C. Kao and W. W. Repko,
Phys. Rev. D {\bf 59}, 013005 (1998).
\bibitem{dkr} D. A. Dicus, K. Kovner and W. W. Repko, Phys. Rev. D {\bf 62},
053013 (2000).
\bibitem{p93} See Chapter 6 of P. J. E. Peebles,
{\it Principles of Physical Cosmology}
(Princeton University Press, Princeton, NJ, 1993).
\bibitem{w72} Strickly speaking, the coefficient 2 in this equation
should be replaced by a factor less than 1, since the temperature
at which we are interested is of the order $10^{13}$ K.
For a discussion, see Chapter 15 of
S. Weinberg, {\it Gravitation and Cosmology}
(John Wiley \& Sons, NY, 1972).
\end{thebibliography}
\end{document}